\documentclass[runningheads]{llncs}
\usepackage{graphicx}
\usepackage{arydshln}
\usepackage{color}
\usepackage{multirow}
\usepackage{amsmath}

\usepackage{hyperref}
\begin{document}
\title{Taxonomy of Mathematical Plagiarism}

\author{Ankit Satpute\inst{1,2}\orcidID{0000-0003-3219-026X} \and
Andr\'{e} Greiner-Petter\inst{2}\orcidID{0000-0002-5828-5497} \and
Noah Gießing\inst{1}\orcidID{0009-0006-5268-2519} \and
Isabel Beckenbach\inst{1}\orcidID{0000-0001-6691-7362}\and
Moritz Schubotz\inst{1}\orcidID{0000-0001-7141-4997}\and
Olaf Teschke\inst{1}\orcidID{0009-0003-4089-9647}\and
Akiko Aizawa\inst{3}\orcidID{0000-0001-6544-5076}\and
Bela Gipp\inst{2}\orcidID{0000-0001-6522-3019}}
\authorrunning{Satpute et al.}
\institute{
FIZ Karlsruhe Leibniz Institute for Information Infrastructure, Berlin, Germany
\email{FirstName.LastName@fiz-karlsruhe.de}\\
\and
Georg August University of Göttingen, Göttingen, Germany\\
\and 
National Institute of Informatics, Tokyo, Japan}
\maketitle              % typeset the header of the contribution

\begin{abstract}
Plagiarism is a pressing concern, even more so with the availability of large language models.
Existing plagiarism detection systems reliably find copied and moderately reworded text but fail for idea plagiarism, especially in mathematical science, which heavily uses formal mathematical notation.
We make two contributions.
First, we establish a taxonomy of mathematical content reuse by annotating potentially plagiarised 122 scientific document pairs.
Second, we analyze the best-performing approaches to detect plagiarism and mathematical content similarity on the newly established taxonomy.
We found that the best-performing methods for plagiarism and math content similarity achieve an overall detection score (PlagDet) of 0.06 and 0.16, respectively.
The best-performing methods failed to detect most cases from all seven newly established math similarity types.
Outlined contributions will benefit research in plagiarism detection systems, recommender systems, question-answering systems, and search engines.
We make our experiment's code and annotated dataset available to the community: 
\url{https://github.com/gipplab/Taxonomy-of-Mathematical-Plagiarism}.

\end{abstract}

\section{Introduction}

Plagiarism is “the use of ideas, concepts, words, or structures without appropriately acknowledging the source in settings expecting originality”~\cite[p.~5]{fishman2009we}.
Plagiarism is a pressing concern as it wastes peer reviewers' time, funding an unoriginal work and depriving original authors of the benefits~\cite{plagDeterrance23,plagEffects2011}.
Large Language Models (LLMs) make plagiarism easier due to produced unreferenced content~\cite{meyer2023chatgpt}.
Most plagiarism detection systems (PDS) have addressed identifying unoriginal text~\cite{homoglyphPlag,lovepreet2019survey} and, to a lesser extent, unoriginal non-textual content such as formulae or images~\cite{Foltynek2019}.
Research on plagiarized mathematical content is incipient, but prior studies have highlighted publications with questionable reuse of math content, leading to retractions~\cite{normanCitplag,Schubotz2019}.

A common approach~\cite{MathIRsurvey2016} to find similar mathematical content is determining whether or not two mathematical formulae are identical.
Due to the complex nature of math and possible underlying assumptions not stated explicitly, it is difficult to judge semantic similarities between expressions since similarity in presentation does not imply a semantic similarity.
The unavailability of an annotated dataset of similar mathematical content in a machine-processable format like LaTeX or MathML currently hinders progress in detecting human-modified similar mathematical content.
For tasks like paraphrase identification (PI), established corpora like ETPC~\cite{kovatchev2018etpc} provide a taxonomy of paraphrasing types.
Establishing these types has been greatly helpful in shaping advanced PI methods~\cite{paraETPCimpro2022,babakov-etal-2022-large} to detect various ways similar text can be represented.
Thus far, a taxonomy of mathematical content similarity in math plagiarised instances hasn't been produced due to the subject's complicated nature~\cite{normanCitplag}, impeding methods to detect human-modified similar math.
Establishing annotated corpora would benefit analysis of the performance of existing math similarity detection systems and the development of new systems.

In this work, we analyze and annotate 122 scientific document pairs from zbMATH Open that experts judged to be potential cases of plagiarism.
We do not claim that the cases are plagiarized. 
The final decision of whether they are plagiarized is up to the field experts.
We establish a taxonomy of mathematical content similarity by identifying a type of reuse and classifying it into a set of rules that describes the reuse.
It has great potential to accelerate LLMs, enabling them to identify similar structural and semantic math contents.

\section{Related work}

The datasets of the workshop series \textit{Plagiarism Analysis, Authorship Identification, and Near-Duplicate Detection} (PAN)~\cite{SteinKS07} are frequently utilized to develop and evaluate PDS.
The PAN datasets consist of artificially created plagiarism, e.g., by randomly removing, inserting, or replacing words or phrases.
The representativeness of simulated plagiarism in PAN datasets to real plagiarism is unclear~\cite{plagRev2019}, limiting the generalizability of evaluation results from these datasets.

For mathematical content, resources of a similar scale to text reuse are missing.
Only two works analyzed mathematical content in scientific documents to identify plagiarism~\cite{normanCitplag,Meuschke2017b}.
Both studied basic mathematical symbol occurrences and used a small evaluation dataset of 10 document pairs.
Currently, No PDS considers semantic textual and non-textual content similarity~\cite{lovepreet2019survey,Foltynek2019}.
While identifying mathematical plagiarism has barely been studied, most mathematics-related tasks focused on search engines and similarity analysis, such as NTCIR~\cite{NTCIR12_2016} and ARQMath~\cite{mansouri2022advancing}.
In NTCIR-12, given a formulae query, relevant formulae were retrieved from the arXiv collection and Wikipedia articles.
The ArqMath-3~\cite{mansouri2022advancing} competition has released formula pairs from MathStackExchange question-answer pairs.
Formulae from the questions and their answers are considered candidates, and the results submitted by participants are ranked on a scale of 0 (Irrelevant) to 3 (Highly relevant) by student annotators.
Even though both NTCIR and ARQMath contain relevant formulae pairs, they were annotated by human post-detection, and none of the pairs have any information about how the content is modified.

Meuschke et al.~\cite{normanCitplag} have formed categories of similar mathematical content by analyzing 10 document pairs retracted for plagiarism collected from computer science, biology, etc., research areas.
However, most categories focus on presentation rather than semantic manipulation. 
The most comprehensive dataset of real-world plagiarised cases in mathematical academia was extracted from zbMATH Open~\cite{Schubotz2019}.
They analyzed 10 document pairs out of the suspected 446, only visually pointing to similar math in a document under inspection and a potential source document.

\section{Dataset}
This work uses real-world plagiarised cases in mathematical documents from zbMATH Open~\cite{Schubotz2019}.
zbMATH Open is the most comprehensive and longest-running (1868 - present) abstracting and reviewing service in mathematics.
The majority of 4.5 million entries in zbMATH Open are in English, and even if the underlying article is not in English, the review, summary, or abstract is typically written in English.
We use 446 entries from zbMATH Open identified for noticeable content reuse~\cite{Schubotz2019} and analyze each entry to annotate which exact parts of the documents are similar, i.e., reuse cases.
Further, we categorize each reuse case into a type, eventually forming a taxonomy of reuse.
We call these reuse types \textit{obfuscation Operators} since they present obstacles to automatic reuse detection.
\textit{obfuscation Operators} does not imply malicious motivation to any author.

\textbf{Annotation procedure: }
zbMATH Open does not contain full-text documents.
We manually collected the full texts by accessing the publisher or repository hosting them. 
Unfortunately, we could only obtain full texts for 122 pairs (inspected document - potential source document) out of 446 due to inactive full-text hyperlinks present at zbMATH Open, removal of full texts by publishers due to retraction notices, etc.
In most cases, we could find the document in PDF format.
Hence, we used MathPiX~\cite{mathpixMathpixAIpowered} to convert PDF to LaTeX, eventually having access to all math formulae in machine-processable format LaTeX.

\begin{table*}
\begin{tabular}{lclcc}
\hline
Obfuscation & Abbre- & Short & \multicolumn{2}{c}{\underline{Cases count}}\\
Operator & viation &  Explanation & Comb. & Uniq. \\
\hline
Paraphrasing & P & Text or Math rewording & 1354 & 133\\
\hdashline
Insertions and & \multirow{2}{*}{ID} & Insert or delete & \multirow{2}{*}{766} & \multirow{2}{*}{21}\\
Deletions &  & text/math expressions & & \\
\hdashline
\multirow{2}{*}{Substitutions} & \multirow{2}{*}{S} & Substitute a term/expression & \multirow{2}{*}{94} & \multirow{2}{*}{2}\\
& & or replace by cross reference & &\\
\hdashline
Text to Math/ & \multirow{2}{*}{TMMT} & Math formulae explained & \multirow{2}{*}{164} & \multirow{2}{*}{1}\\
Math to Text & & with text words or vice versa & &\\ 
\hdashline
Different & \multirow{2}{*}{DP} & Objects with the same meaning& \multirow{2}{*}{733} & \multirow{2}{*}{20}\\
Presentation & & but different presentations & & \\
\hdashline
Formula & \multirow{2}{*}{FM} & Substituting formulae with a different repr.& \multirow{2}{*}{228} & \multirow{2}{*}{2}\\
Manipulation & &  of semantic content through transformations & &\\
\hdashline
Variation of & \multirow{2}{*}{VS} & Semantics different but argumentative & \multirow{2}{*}{959} & \multirow{2}{*}{57}\\
Subject & & structure similar & & \\
\hline
\end{tabular}
\footnotesize
\caption{Overview of all obfuscation operators and count of annotated cases per obfuscation operator. Comb: cases occurring with other obfuscation. Uniq: cases occurring with only one particular obfuscation.}
\label{table:obfuscOver}
\end{table*}

A group of 7 field experts with a postgraduate degree in Mathematics and 6 having Ph.D. annotated similar mathematical contents.
They used the publicly available annotation tool TEIMMA~\cite{satpute2023teimma} for annotating reuse.
They regarded content from an inspected document as an instance of content reuse if it overlaps significantly with the content from the source document.
They assigned \textit{obfuscation operators} to each case, depicting the possible way in which content from the source document was modified.
In most annotations, multiple obfuscation operators were found simultaneously.
In the following, we describe each obfuscation operator formed by annotators.

\subsection{Proposed Taxonomy of Mathematical content similarity:}
Table~\ref{table:obfuscOver} presents a summary of all obfuscations.
The definitions and rules for each obfuscation operator were fixed gradually, with frequent discussions with experts with experience reviewing documents for zbMATH Open.

A taxonomy needs a precise vocabulary for categories. 
Terms like formulae and expressions are often used interchangeably~\cite{studMathDefUse2004,unifiedMathdef2023}. 
Thus, we initially set definitions for obfuscation operator terms.
\textbf{Formula: }A combination of mathematical symbols formed in accordance with the rules of mathematical syntax. 
\textbf{Maximal Expressions: }A well-formed expression inside a mathematical formula that cannot be extended further without becoming a mathematical statement. 
For example, $a+2$ and $0$ in mathematical statement $a+2=0$. 
Non-maximal expressions will be called subexpressions. 
For the above example, subexpressions would be $a$, $2$, and $0$. 
\textbf{Synonymous: } Two expressions of almost identical semantics are synonymous when representing the same mathematical object. For example, $1$ and $1.0$ are synonymous.

\subsubsection{Obfuscation Operators: }
\noindent\textbf{1. Paraphrasing (P)}\label{parparobf}: Paraphrases convey the same meaning while using different words or sentence structures~\cite{paraphraseDefin}.
We extend the definition to mathematical content paraphrases by placing two restrictions. 
First, a mathematical content paraphrase requires that all maximal expressions from the source document must be preserved without changes.
In addition, relators (greater than, less than, etc.) may be changed according to necessity.
\newline \textit{Example Case:} \{ \textit{Src :}The proof reduces to showing that $x\leq 1$. \}
\{ \textit{Insp : }It is sufficient to prove that $1\geq x$.\}

\noindent\textbf{2. Insertions and Deletions (ID)}
A pair of text passages are subject to Insertions if a similar passage adds redundant or insubstantial content. 
The definition for deletions is similar, only reversed: insubstantial content from the source passage is deleted.
\newline \textit{Example Case:} Notice the middle expression in \textit{Insp : }   
\{ \textit{Src :} \(\langle x+y,x+y \rangle \leq \lVert x \rVert^2 + 2 \lvert \langle x,y \rangle \rvert + \lVert y \rVert^2\) \}
\newline \{ \textit{Insp : } \( \langle x+y,x+y \rangle \ {=\lVert x \rVert^2 + \langle x,y \rangle + \langle y,x \rangle + \lVert y \rVert^2} \\ \leq \lVert x \rVert^2 + 2\lvert\langle x,y \rangle\rvert + \lVert y \rVert^2. \) \}

\noindent\textbf{3. Substitutions (S)}
We consider a reuse case as a substitution when either of the following occurs:
1. In formulae: replacement of expression $e$ in the source by another expression $f$, which is related to $e$ by a formula $eRf$ for some relator $R$, or vice versa.
2. In-text: replacement of textual content in the source by cross-reference to an earlier passage with equivalent content, or vice versa.
\newline \textit{Example Case:} Content in \textit{Insp} eventually leads to \textit{Src} by substitutions. 
\{\textit{Src : }$x+3=y-2$ \}
\{\textit{Insp : }$A(x)=B(y),$ where $A(x)=x+3$ and $B(y)=y-2$. \}

\noindent\textbf{4. Text to Math/Math to Text (TMMT)}
In the Text to Math obfuscation operator, mathematical content from the source document expressed in text is substituted by mathematical notations. 
Similarly, Math to Text describes the reverse obfuscation operation.
\newline \textit{Example Case:}
\{ \textit{Src : }The force $F$ acting on a body equals the product of its mass $m$ and acceleration $a$. \}
\{ \textit{Insp : $F=ma.$} \}

\noindent\textbf{5. Different Presentation (DP)}
This operator allows two notational changes.
First, an expression is replaced by a synonymous expression.
Second, if two expressions are single objects, all root and child objects in one expression are matched to synonymous operators and objects in another.
\newline \textit{Example Case:} A simple change of variable names.
\{\textit{Src :}$f(x)$ \}
\{\textit{Insp :}$g(x)$ \}

\noindent\textbf{6. Formula Manipulation (FM)}
Formula Manipulation refers to substituting formulae with different representations of their semantic content obtained through transforming expressions using algebraic identities or statements using logical equivalences or implications. 
In contrast to the \textit{substitutions} operator, the equivalence of transformed expressions is not announced elsewhere in the document and has to be logically deduced.
Examples of methods to transform expressions could be expanding a single term into multiple terms using rules such as distributivity, associativity, etc. 
\newline \textit{Example Case:} The following example involves applying coordinate changes, e.g., expressing a complex number from \textit{Src} in polar coordinates in \textit{Insp}.
\{ \textit{Src : }$z = a+bi$ \}
\{ \textit{Insp : }$z=re^{i\phi}$ \}

\noindent\textbf{7. Variation of Subject (VS)}
In the Variation of the Subject obfuscation operator, the semantic content of a passage after the application is strictly different from the original passage, but the argumentative structure remains similar.
\newline \textit{Example Case:} This is a non-mathematical (Included due to space limitation. Please refer to the repository\footnote{\url{https://github.com/gipplab/Taxonomy-of-Mathematical-Plagiarism}} for a real example) example that illustrates how content might be modified in several places to accommodate the new setting. 
\{ \textit{Src : }I went to [Paris] for a few days and visited the (Eiffel Tower).\}
\{ \textit{Insp : }I went to [Rome] for a few days and visited the (Colosseum).\}

\textbf{Annotation verification by an expert: }Since similar mathematical content annotations were done in a single instance by field experts, verifying if their annotations are consistent is essential.
We provided 10\% document pairs to an independent expert with a Ph.D. in mathematics to do similar annotations.
Creating manual annotations is time-consuming, and it is hard to find experts in multiple disciplines to produce adequate annotations.
We provided obfuscation categories defined above to assign to the selected spans.
The independent expert annotated the document pairs using TEIMMA~\cite{satpute2023teimma}. 
We use Jaccard similarity~\cite{bank2008calculating} to find the overlap between the original annotations and the independent expert's annotations, i.e., for each document pair.
On average, there was 80.01\% token overlap (both text and math).
Similarly, for exact \textit{Case Type} (Whether reuse contains text, math, or both) and exact \textit{Obfuscation Type} matches, there was 95.21\% and 74.23\% overlap, respectively.
For obfuscation types, we further calculate inter-annotator agreement using the kappa value and achieve a score of 0.39, indicating a fair agreement.
We understand that agreeing upon the same obfuscations is difficult as it is subjective.

\section{Evaluation}
\begin{table}
\centering
\begin{tabular}{ccccccccccccccccccc}
\hline
Obfusc.$\rightarrow$ & \multicolumn{3}{c}{All} & & & \multicolumn{3}{c}{P(78.55\%)} & & & \multicolumn{3}{c}{ID(44.86\%)} & & & \multicolumn{3}{c}{S(5.39\%)}\\
\hdashline
Method $\downarrow$ & F1 & G & PD & & & F1 & G & PD & & & F1 & G & PD & & & F1 & G & PD\\
\hline
LCIS & 0.00 & \textbf{1.00} & 0.00 & & & 0.00 & \textbf{1.00} & 0.00 & & & 0.00 & \textbf{1.00} & 0.00 & & & 0.00 & \textbf{1.00} & 0.00\\ 
GIT & 0.05 & 1.28 & 0.04 & & & 0.05 & 1.35 & 0.04 & & & 0.02 & 1.38 & 0.02 & & & 0.00 & \textbf{1.00} & 0.00\\ 
AdaPlag & 0.08 & 1.34 & 0.06 & & & 0.08 & 1.39 & 0.06 & & & \textbf{0.09} & 1.29 & \textbf{0.07} & & & \textbf{0.10} & 1.36 & \textbf{0.08}\\
MABOWDOR & \textbf{0.17} & 1.10 & \textbf{0.16} & & & \textbf{0.13} & 1.27 & \textbf{0.11} & & & 0.04 & 1.02 & 0.04 & & & 0.00 & \textbf{1.00} & 0.00\\  
\hline
\hline
Obfusc.$\rightarrow$ & \multicolumn{3}{c}{TMMT(9.50\%)} & & & \multicolumn{3}{c}{DP(45.89\%)} & & & \multicolumn{3}{c}{FM(12.92\%)} & & & \multicolumn{3}{c}{VS(55.30\%)}\\
\hdashline
Method$\downarrow$ & F1 & G & PD & & & F1 & G & PD & & & F1 & G & PD & & & F1 & G & PD\\
\hline
LCIS & 0.00 & \textbf{1.00} & 0.00 & & & 0.00 & \textbf{1.00} & 0.00 & & & 0.00 & \textbf{1.00} & 0.00 & & & 0.00 & \textbf{1.00} & 0.00\\ 
GIT & 0.00 & 2.00 & 0.00 & & & 0.02 & 2.00 & 0.01 & & & 0.01 & 2.00 & 0.01 & & & 0.02 & 1.42 & 0.02\\ 
AdaPlag & \textbf{0.07} & 1.38 & \textbf{0.05} & & & \textbf{0.08} & 1.30 & \textbf{0.07} & & & \textbf{0.09} & 1.33 & \textbf{0.07} & & & 0.07 & 1.17 & 0.06\\
MABOWDOR & 0.00 &  \textbf{1.00} & 0.00 & & & 0.07 & 1.24 & 0.06 & & & 0.02 & 1.09 & 0.02 & & & \textbf{0.09} & 1.09 & \textbf{0.09}\\
\hline
\end{tabular}
\caption{Evaluation of plagiarism detection and Math Content Similarity methods on the new dataset. Values in parentheses indicate the percentage of cases of all cases in which a particular obfuscation is present. (Lower G and higher F1 \& PD represents better detection)}
\label{table:evalMathAll}
\end{table}

We use four detection approaches.
The first two are the Longest Common Subsequence of Identifier (LCIS) and Greedy Identifier Tiles (GIT) presented in the only prior work on math plagiarism detection using mathematical content by Meuschke et al.~\cite{normanCitplag}.
Third, AdaPlag, the winning approach of the PAN plagiarism competition~\cite{sanchez2015adaptive}.
Fourth, a Math-Aware Best of-Worlds Domain Optimized Retriever (MABOWDOR)~\cite{mabowdor2023}, which is the best-performing math content similarity approach on the ARQMath dataset~\cite{mansouri2022advancing}.

We create a test collection to represent a real-world retrieval scenario.
zbMATH Open does not contain full texts, but around 464K entries correspond to documents from the arXiv.
We obtained LaTeX sources of these 464K entries from the arXMLiv dataset~\cite{SML:arXMLiv:2020}.
We include the corresponding source documents of all annotated 122 inspected documents in the test collection and remove inspected documents from the test collection to avoid exact matches.
For each of the four methods, we query our test collection of ~464K documents by each of the 122 inspected documents.
We combine two steps (source retrieval and detailed comparison) of typical PDS into one, i.e., direct comparison without any hyperparameters tuning.
For evaluation, we use the F1 score, i.e., the harmonic mean of precision and recall), Granularity (G), which determines whether a reuse case was detected as a whole or in several pieces, and PlagDet (PD), which combines recall, precision, and G to allow for ranking, are standard evaluation metrics for plagiarism detection~\cite{Plageval2010}. 
For each of the 122 inspected documents, we take the top 10 most similar documents out of the test collection and calculate evaluation scores.
Table~\ref{table:evalMathAll} shows the evaluation of approaches on the newly constructed dataset.
Results indicate that most of the cases from all obfuscation operators remain undetected by existing methods.
Improvements can likely be made by hyperparameter tuning.

\section{Conclusion}

In this work, we established a novel taxonomy of math content similarity.
We formed 7 math content similarity types by annotating 122 document pairs identified for potential plagiarism.
We analyzed the performance of the best-performing plagiarism detection methods and math content similarity in detecting cases from newly established taxonomy. 
It was found that the current best-performing methods do not detect most human modifications on math content (Overall PlagDet score of 0.06 and 0.16 for the best-performing plagiarism and math content similarity methods, respectively).
The dataset curated in this work and the code of the experiments are publicly available to aid future research on math content similarity detection.
A resource and experiments presented in this paper set up a base to develop advanced detection techniques capable of identifying modified math content reuse.
This work will accelerate research identifying concealed plagiarism instances in academia and research involving mathematical content similarity.

\section*{Acknowledgements}

This work was funded by the Deutsche Forschungsgemeinschaft (DFG, German Research Foundation) - 437179652 and 460135501, the Deutscher Akademischer Austauschdienst (DAAD, German Academic Exchange Service - 57515245), and the Lower Saxony Ministry of Science and Culture and the VW Foundation.

\bibliographystyle{splncs04}
\bibliography{main}
\end{document}